\documentclass[12pt,preprint]{aastex}

\shorttitle{Temperature Structure of the Quiet Corona}
\shortauthors{Brooks et al.}
\begin{document}

\title{{\it Hinode}/Extreme-Ultraviolet Imaging Spectrometer Observations of the Temperature Structure of the Quiet Corona}
\author{David H. Brooks \altaffilmark{1,2,3}, Harry P. Warren \altaffilmark{1}, David R. Williams \altaffilmark{3,4}, Tetsuya Watanabe \altaffilmark{5}}
\altaffiltext{1}{Space Science Division,  Code 7673, Naval Research Laboratory, Washington, DC 20375}                       
\altaffiltext{2}{George Mason University, 4400 University Drive, Fairfax, VA 22020}                                  
\altaffiltext{3}{Present address: Hinode Team, ISAS/JAXA, 3-1-1 Yoshinodai, Sagamihara, Kanagawa 229-8510, Japan}
\altaffiltext{4}{Mullard Space Science Laboratory, University College London, Holmbury St Mary,  Dorking, Surrey, RH5 6NT, UK}     
\altaffiltext{5}{National Astronomical Observatory of Japan, Osawa, Mitaka, Tokyo 181-8588, JAPAN}     
\email{dhbrooks@ssd5.nrl.navy.mil}

\begin{abstract}
We present a Differential Emission Measure (DEM) analysis of the quiet solar corona on disk using data obtained
by the Extreme-ultraviolet Imaging Spectrometer (EIS) on {\it Hinode}. We show that the expected quiet Sun
DEM distribution can be recovered from judiciously selected lines, and that their average intensities can be reproduced
to within 30\%. We present a subset of these selected lines spanning the temperature range $\log$ T = 5.6 to 6.4 K 
that can be used to derive the DEM distribution reliably, including a subset of Iron lines that can be used to
derive the DEM distribution free of the possibility of uncertainties in the elemental abundances. The 
subset can be used without the need for extensive measurements and the observed intensities can be
reproduced to within the estimated uncertainty in the pre-launch calibration of EIS. 
Furthermore, using this
subset, we also demonstrate that the quiet coronal DEM distribution can be recovered on size scales down
to the spatial resolution of the instrument (1$''$ pixels).          
The subset will therefore be useful
for studies of small-scale 
spatial inhomogeneities in the coronal temperature structure, for example, in addition to studies 
requiring multiple DEM derivations in space or time.
We apply the subset to 45 quiet Sun datasets taken in the period 2007 January to April, 
and show that although the absolute magnitude of the coronal DEM may scale with the amount
of released energy, the shape of the distribution is very similar up to at least $\log$ T $\sim$ 6.2 K in all 
cases. This result is consistent with the view that
the {\it shape} of the quiet Sun DEM is mainly a function of the radiating and conducting
properties of the plasma and is fairly insensitive to the location and rate of energy deposition.
This {\it universal} DEM may be sensitive to other factors such as loop
geometry, flows, and the heating mechanism, 
but if so they cannot vary significantly from quiet Sun region to region.
\end{abstract}

\bibliographystyle{/home/brooks/latex/dhb_bib/apj}

\section{Introduction}
Since the advent of space based solar EUV and X-ray observations, spectroscopic diagnostic
techniques have been used to measure solar plasma properties such as the electron density,
velocity, chemical composition, or coronal temperature structure. These quantities are 
useful inputs to physical models and provide confirmation of theoretical predictions. 

The differential emission measure (DEM) technique is widely employed for studying the distribution
of material as a function of temperature. It is also used for validation of atomic data, spectral line identification,
measurements of elemental abundances, instrument radiometric calibration, or to compare with
theoretical models. All these methods, however, rely on a number of physical assumptions
that may not be realistic for the target under investigation. It is important, therefore, to
understand the limitations of the diagnostic capabilities of the techniques so that the measurements can be interpreted correctly.

In recent years, orbiting instrument capabilities have reached the point where spatially resolved
structures in the solar atmosphere can be examined spectroscopically. Variations of the DEM and
emission measure techniques have
been used to try to shed light on the coronal loop controversy: are loops isothermal? 
(\citealt{lenz_etal1999,aschwanden_etal1999,aschwanden_etal2000}) or multi-thermal? (\citealt{schmelz_etal2001,warren_etal2008a}), 
to study the validity of complex multiple peaked DEM
distributions in active regions (\citealt{brosius_etal1996,lanzafame_etal2002}),
and to study spatial inhomogeneities in the quiet
coronal DEM distribution \citep{lanzafame_etal2005}. Most recently, DEM analysis has been applied to observations
from the Extreme ultraviolet Imaging Spectrometer \citep[EIS,][]{culhane_etal2007}
on {\it Hinode} \citep{kosugi_etal2007}.

Previous studies of small-scale spatial inhomogeneities in the quiet coronal DEM distribution
have suggested that the shape of the DEM is very similar up to $\log$ T$_e$ = 6.1 K
when it is derived from intensities 
averaged over large enough areas \citep{lanzafame_etal2005}. 
If confirmed, this is potentially an important result because it provides a further constraint on the heating.
Our interest here is to extend this analysis 
of small-scale spatial variations
in the quiet coronal DEM to multiple regions observed at different locations and different times and therefore examine
in detail whether this result is true in general.

For ambitious studies that require derivation of the DEM distribution multiple
times, this can be time consuming when many lines are included in the analysis. For the EIS data, 276 spectral lines
were identified in the 171--212\,\AA\, and 245--291\,\AA\, wavelength ranges by \citet{brown_etal2008} and each line 
would need careful fitting prior to a DEM study. 
It is of interest therefore to produce a 
subset of spectral lines that can be used to reliably derive the DEM in multiple locations
or at many different times. 

EIS also has the capability  
to derive the DEM distribution using spectral lines of Fe ions only, thus allowing an abundance
uncertainty free analysis \citep{watanabe_etal2007}. 
In addition, Fe ions have the advantage
that they relax into ionization equilibrium faster than ions of other species \citep{lanzafame_etal2002}
which could be helpful for probing the assumptions underlying the technique. 
It is also of interest, therefore, to emphasize Fe when selecting reduced subsets of reliable spectral lines.

It is well known, however, that current atomic modeling of this complex atom is 
insufficiently advanced compared to that of lighter species \citep{summers_etal2006},
and recent analysis of EIS data has indicated significant uncertainty in the ionization
equilibrium calculations for Fe \citep{young_etal2007a}
and modeling of the weaker emission lines.
In order to achieve our aims, we need to have confidence in the atomic data used for the
analysis, and to examine the spectra in detail to understand which lines 
are the best candidates for detailed analysis. In addition, one needs to have 
confidence that the photometric calibration performed in the laboratory can be
used reliably on orbit. 

In this paper, we first examine a spectral atlas prepared from EIS observations of a very quiet
region observed on 2007 January 30. A quiet coronal region was chosen so that comparison could
be made to previous coronal DEM distributions derived from quiet Sun data. 
Here we compare with the DEM distribution obtained
by \citet{brooks&warren_2006} using
Solar and Heliospheric Observatory
\citep[SOHO,][]{domingo_etal1995} Coronal Diagnostic Spectrometer \citep[CDS,][]{harrison_etal1995}
observations. We derive the DEM distribution using the most reliable lines
and apply that solution to a larger linelist to identify problematic issues with line
fitting, atomic data etc. We then produce a subset of lines, with one recommended line
at each temperature, that can be used to recover
the same DEM solution. 
This subset can be used without the need for extensive measurements 
and contains a series of lines from Fe ions that can be used to derive the DEM removing the
possibility of uncertainties due to elemental abundances. 
We also demonstrate here that EIS is an ideal
instrument for studying inhomogeneities in the DEM distribution because 
its sensitivity allows us to derive the same coronal DEM 
on very small spatial scales (down to a single EIS 1$''$ pixel). 

We then apply our best-case EIS DEM solution to 45 full-CCD quiet Sun datasets and explore
the possibility 
that it can reproduce the observed intensities of the subset of lines in each observation.
We find that in the vast majority of the cases the DEM is very similar up to at
least $\log$ T$_e$ = 6.15 K with only minor adjustments to the shape and some re-scaling
of the magnitude. 

In \S \ref{oadr} we describe the observations studied and data reduction techniques used.
In \S \ref{ad} we reference the atomic data adopted. In \S \ref{dem} we describe the 
DEM method and specifics of our analysis. In \S \ref{lss} we discuss our analysis 
strategy. In \S \ref{res} we present our results and in \S \ref{conc} we summarize
the paper and discuss the implications of our findings.

\section{Observations and Data Reduction}
\label{oadr}
{\it Hinode} made observations of a quiet region close to disk center on 2007 January 30. Figure \ref{fig1}
shows an EIT 195\,\AA\, full-disk image taken at 13:13:40UT with the EIS study field-of-view overlaid. 
The FOV is 128$'' \times$128$''$ and was built up by stepping the 1$''$ slit from solar West to East
over a 3 1/2 hour period from 14:34:40 to 18:00:00UT.
The study takes full spectra on both the EIS detectors from 171-212\,\AA\, and 245-291\,\AA. The exposure
time was 90s and the study acronym is HPW001\_FULLCCD\_RAST. The data were taken during the daily long SAA (South Atlantic
Anomaly) free period 
around the {\it Hinode} orbit. 

Figure \ref{fig2} shows example images formed from Gaussian line fits to the EIS data. Examples using
\ion{Si}{7} 275.352\,\AA, \ion{Fe}{10} 185.213\,\AA, \ion{Fe}{12} 195.119\,\AA, and \ion{Fe}{15} 284.160\,\AA\,
are shown. These lines are formed in the temperature range $\log$ T$_e$ = 5.7 to 6.35 K.  Both Figures
\ref{fig1} and \ref{fig2} indicate that the region of quiet corona selected is indeed free of any
significant activity.

The EIS data require removal of the CCD dark current, cleaning for cosmic-ray strikes on the CCD, and
processing of the data to take account of hot, warm, and dusty pixels. In addition, the radiometric
calibration needs to be applied to convert the data from photon events to physical units. In this 
work we have applied these corrections and calibrations using the EIS software routine {\it EIS\_PREP} that is available in SolarSoft.
Average line profiles over the full 128$''$ in the X- and Y- directions were constructed to improve the signal-to-noise
and provide a spectral atlas for analysis. 

The EIS CCDs were slightly offset from each other by 1--2 pixels in -X prior to an adjustment to the slit assembly position
on 2008, August 24. In addition, there is an offset between the CCDs of $\sim$ 18 pixels in -Y (wavelength
dependent). No major difference is expected to be introduced, however, by
ignoring these offsets for the spectral atlas: on disk quiet Sun intensities typically do not vary significantly 
when averaged over large spatial areas. Using the Fe lines from the final subset (see \S \ref{demsubset}),
we verified that in these observations the detector offsets typically
introduce a variation of less than 15\% to the averaged line intensities. 
The offsets are clearly important for velocity analysis and accurate measurements in small areas, so we have 
taken the appropriate values into account 
for the DEM analysis of the single pixel measurements (see \S \ref{demsinglepix}). 

\section{Atomic Data}
\label{ad}
In this work, we use the CHIANTI database version 5.2 (\citealt{dere_etal1997,landi_etal2006}). Details
of the sources of the 
electron collisional excitation, deexcitation, and spontaneous
radiative decay rates, are given in the database and accompanying papers. The zero density ion fractional abundances
of \citet{mazzotta_etal1998} are used throughout the paper, together with the coronal elemental
abundances of \citet{feldman_etal1992}.

\section{Differential Emission Measure}
\label{dem}
Assuming a constant electron pressure as the basis of a relationship between electron temperature
and density, an optically thin emission line intensity arising from a transition between two
atomic levels can be written as
\begin{equation}
I_{n l } = A(Z) \int_{T_e} G_{n l} (T_e, N_e) \phi (T_e) dT_e
\end{equation}
where $n$ and $l$ denote the levels, $A(Z)$ is the elemental abundance of the atomic
species, $N_e$ and $T_e$ are the electron density and temperature, $\phi (T_e)$ is the emission
measure differential in temperature, and $G (T_e, N_e)$ is the contribution function. The latter
is a function of temperature and density but is calculated in this work at a specific electron pressure
estimated from a density sensitive line ratio in \S \ref{cep}. The results of that measurement
justify the neglect of density effects for the very quiet region studied in this work.

\subsection{Method}                
\label{m}
We represent the emission measure curve with a series of spline knots as in \citet{warren_2005}
and \citet{brooks&warren_2006}.
These knots can be interactively moved to control the smoothness of the emission measure distribution
and this has been found to be helpful in the past for representing the rapid fall-off of 
emission measure at high temperatures in the quiet corona. The values of the knots are determined
from a $\chi^2$ minimization of the differences between the measured intensities and those calculated
from the DEM distribution.

\subsection{Coronal Electron Pressure}                
\label{cep}
As discussed above, we need to make an electron density measurement using line ratios in order to
calculate the contribution functions that are needed for the DEM analysis. The EIS wavelength 
range contains a number of excellent density sensitive line ratios and some of these have been
assessed in detail by \citet{young_etal2009}. Here we use the \ion{Fe}{12} 186.8\,\AA\, to \ion{Fe}{12} 195.119\,\AA\,
ratio that is sensitive in the density range $\log$ N$_e$ = 8--12 cm$^{-3}$. 

\citet{young_etal2007a} note that the \ion{Fe}{12} 186.8\,\AA\, line is a self-blend of lines at 186.88\,\AA\, and 186.856\,\AA\,
and is also blended with \ion{S}{11} 186.839\,\AA. They also note that the contribution from the \ion{S}{11} 186.839\,\AA\,
line can be estimated by taking the ratio of that line to the \ion{S}{11} 191.264\,\AA\, line. This ratio
has a fixed value of 0.2. Furthermore, the {\ion{Fe}{12} 195.119\,\AA\, line has been found to be broader
in EIS spectra than the \ion{Fe}{12} 193.515\,\AA\, line \citep{young_etal2007a}, indicating a possible \ion{Fe}{12} blend at 195.18\,\AA\,
as suggested by
\citet{delzanna&mason_2005}. If present, this blend should only be significant at high densities \citep{delzanna&mason_2005}
and should only contribute $\sim$10\% to the main feature
at densities of $\log$ N$_e$ = 10 cm$^{-3}$ \citep{young_etal2007a}. 

We measured the intensity of the \ion{S}{11} 191.264\,\AA\, line in order to estimate the contribution
of the \ion{S}{11} 186.839\,\AA\, line to the component identified as \ion{Fe}{12} 186.8\,\AA. We then
calculated the spread in the diagnostic line ratio values that would be obtained if we included or
excluded this blend and also included or excluded a 10\% modification to the \ion{Fe}{12} 195.119\,\AA\,
line intensity arising from the possible blend at 195.18\,\AA.
The measured ratios spread from 0.12--0.15 and these values translate to a spread
in electron density of $\log$ N$_e$ = 8.4--8.5 cm$^{-3}$ or 
electron pressure of $\log$ P$_e$ = 14.5--14.65 cm$^{-3}$ K. 
These values are in good 
agreement with the quiet coronal electron pressure calculated from \ion{Si}{10} and \ion{Si}{11} line ratios 
by \citet{brooks&warren_2006} and also with previous measurements by \citet{young_2005} and
\citet{warren_2005}. Thus the blends do not appear to significantly affect our measurements with 
this ratio and we therefore adopt a value of $\log$ P$_e$ = 14.5 cm$^{-3}$ K for calculating the 
contribution functions.

\subsection{Line Selection Strategy}
\label{lss}
We used the comprehensive spectral line identification study of \citet{brown_etal2008} to identify
276 lines in our quiet Sun dataset. \citet{brown_etal2008} state that those lines with greater than
20DN per 60s exposure have good precision in their Gaussian fitting. Therefore, we excluded lines 
with lower count rates from our analysis with the exception of the following four:
\ion{Ne}{5} 184.73\,\AA, \ion{O}{4} 279.631\,\AA, \ion{O}{4} 279.933\,\AA, and \ion{O}{6} 183.937\,\AA.
These lines were included in the hope of stabilizing the DEM distribution at lower temperatures: 
the vast majority of the spectral lines in the EIS wavelength range are formed at upper transition
region and coronal temperatures. 
In addition, referring to
the first results analysis of the EIS spectra of \citet{young_etal2007a} and again to \citet{brown_etal2008}
we excluded lines that had been identified as blended in either study. Thus the following 19 lines
were excluded: 
\ion{Ar}{11} 184.501\,\AA, 
\ion{Fe}{12} 186.880\,\AA, 
\ion{Ar}{14} 187.954\,\AA, 
\ion{Fe}{11} 189.125\,\AA, 
\ion{Fe}{11} 189.940\,\AA, 
\ion{Fe}{11} 190.886\,\AA, 
\ion{ S}{11} 191.260\,\AA, 
\ion{Fe}{14} 192.629\,\AA, 
\ion{Fe}{11} 192.811\,\AA, 
\ion{Fe}{11} 192.913\,\AA, 
\ion{Fe}{14} 193.270\,\AA, 
\ion{Fe}{12} 196.645\,\AA, 
\ion{Fe}{11} 198.550\,\AA, 
\ion{Fe}{13} 201.122\,\AA, 
\ion{Fe}{10} 256.434\,\AA, 
\ion{Fe}{12} 259.491\,\AA, 
\ion{Si}{ 7} 276.863\,\AA, 
\ion{Mg}{ 7} 277.021\,\AA, and
\ion{Mg}{ 7} 278.410\,\AA. 
Finally, the following three \ion{Fe}{9} lines, recently proposed as identifications
by \citet{young_2009}, were added to our line list in the hope of filling in the temperature range at the base of the corona:
\ion{Fe}{9} 188.485\,\AA, \ion{Fe}{9} 189.940\,\AA, and \ion{Fe}{9} 197.858\,\AA. The 188.485\,\AA\, and 189.940\,\AA\,
features were identified with lines of \ion{Fe}{12} and \ion{Fe}{11}, respectively, by \citet{brown_etal2008}.
Each of the remaining 77 lines were fitted using 
the \citet{brown_etal2008} analysis as a reference guide for the
positions of spectral features in the spectrum.

The spectral ranges selected were fitted with
multiple Gaussians plus a polynomial background using a line fitting software package developed
by the authors that utilizes the IDL routine {\it MPFIT} \citep{markwardt_2009}. 

After completing this process we further inspected the reliability of the line fits that were made.
Our intention was to try to derive a DEM distribution that could reproduce the observed intensities
to an accuracy level that was comparable to that obtained for previous spectrometers.
Therefore, we
decided to initially select only those lines, the intensities of which were measured to extremely high precision. After
deriving the DEM distribution for these lines we would then apply the solution to 
the complete set of spectral lines. 
In addition,
the following lines were excluded because no suitable atomic data were available:
\ion{Ni}{11} 186.978\,\AA,
\ion{Fe}{11} 192.014\,\AA,
\ion{Fe}{7} 196.054\,\AA,
\ion{Ni}{14} 196.226\,\AA,
\ion{Ni}{14} 200.692\,\AA,
\ion{Fe}{13} 202.424\,\AA, the following two lines were excluded because the lines were too weak
to fit reliably in our spectrum:
\ion{ S}{13} 256.684\,\AA,
\ion{Fe}{16} 262.983\,\AA, and the
\ion{He}{2} 256.332\,\AA\, line was excluded because it is unlikely to be optically thin.

The final set of 29 lines for our first DEM attempt is given in Table \ref{tab1} and the
DEM results are discussed in \S \ref{res}. 

\section{Results}
\label{res}
Here we discuss the results of our DEM analysis. We used the DEM distribution for the
quiet Sun obtained by \citet{brooks&warren_2006} using CDS data as an initial estimate for the EIS data. We then
modified the shape and magnitude to reproduce as many of the 29 spectral line intensities
as possible to within our hoped for tolerance level of 30\%.

\subsection{Best Case DEM}
\label{resbcd}
Figure \ref{fig3} shows the DEM solution over the temperature interval $\log$ T$_e$ = 5.4--6.6 K.
Five of the spline knots used are shown by the boxes on the plot. The other two are positioned
at $\log$ T$_e$ = 4.5 and 5.05 K, and are outside
the temperature range plotted and the temperature range over which the solution
is well constrained. The shape of the DEM distribution ended up being remarkably similar to that found in the
previous study. As in that study, and \citet{warren_2005},
the DEM distribution in the $\log$ T$_e$ = 5.8--6.3 range requires close spacing of the knots 
to represent the sudden decrease in emission measure that is characteristic of the
usual quiet Sun DEM distribution at high temperatures. The lower panel shows ratios of observed
to DEM predicted intensities for the lines of Fe, S, and Si used in
the analysis: only lines of these atomic species made it into the final set of 29. These
ratios are plotted at the temperature of the peak of the relevant contribution function. 

Table \ref{tab1} gives the detailed observed intensities and numbers from the computations.
The temperature of peak abundance in ionization equilibrium for each line is also given,
though, as already stated, there is evidence that these calculations
and therefore the ion formation temperatures, may need revision \citep{young_etal2007a}.
The wavelengths are taken from the literature values of \citet{brown_etal2008}.
It has been noted before that there is a difficulty in reproducing both the Fe
and Si lines simultaneously with the DEM method. \citet{warren_2005} addressed
this issue by increasing the Si abundance by 30\% in his analysis, while
\citet{brooks&warren_2006} did the same but also increased the Fe abundance
by 10\%. Both these analyses point to a 20--30\% relative modification of the abundances
between Fe and Si. Here we found that a slightly larger relative modification, this time
reducing the Fe abundance by $\sim$ 40\%, brings the Fe and Si lines into
better agreement.

Table \ref{tab1} shows that the majority of the lines (70\%) are reproduced to within
$\sim$30\% or better, which was our stated goal. Only 8 lines (27\%) are reproduced to worse
than $\sim$50\% and possible reasons for this are discussed below. Above $\log$ T$_e$ = 6.1 K,
the distribution is well constrained with 11 of the 13 Fe and Si lines 
reproduced to within $\sim$ 30\%. These results are also comparable to those we obtained 
in our previous study. 
There is an interesting contrast with CDS, however, in that the CDS DEM is well constrained 
in the lower temperature range ($\log$ T$_e$ = 5.1--5.7 K) whereas the EIS DEM is best 
constrained at higher temperatures (above $\log$ T$_e$ = 6.1 K). This is as one 
would expect given the different target wavelength and temperature regimes of the two
instruments, but also indicates that there is greater uncertainty in our understanding of
the DEM distribution between $\log$ T$_e$ = 5.7--6.1 K.

Figure \ref{fig4} shows the \citet{brooks&warren_2006} DEM (dashed line),
and the current DEM derived using EIS data (solid line). 
The curves are plotted over the same temperature range as in Figure \ref{fig3}. In the lower panel, 
the ratio of the two DEMs is also shown.
The DEM curves 
remain within 50\% of each other 
until they fall rapidly at high temperatures (above $\log$ T$_e$ = 6.2 K). This
illustrates the similarity in the shapes of the two DEM curves, despite being derived
from different regions using different instrumentation. 

\subsection{Discrepancies between Observed and Predicted Intensities}
The accuracies of the predicted line intensities can be seen in Table \ref{tab1}. 
The following lines were reproduced to worse than 50\% accuracy:

1. The Cl-like \ion{Fe}{10} lines at 193.715\,\AA, 195.399\,\AA, and 257.262\,\AA\,
are all observed to be stronger than the DEM predicts by varying factors. 

Examining
our fit 
to the 193.715\,\AA\, line in detail we are not convinced that it is perfect because 
the line blends into the wings of the nearby \ion{S}{10} 193.469\,\AA\,
line. We find, however, that further attempts to refine our fit lead to an increased
measured intensity for the 193.715\,\AA\, line which exacerbates the mismatch problem.
We can only reduce the intensity if the line is blended, but there is no obvious candidate. 
The CHIANTI database gives an \ion{Fe}{10} line of uncertain wavelength at 193.663\,\AA\,
but it is not expected to contribute significantly to the feature. The NIST database
gives a \ion{Na}{8} line at 193.558\,\AA\, which could conceivably disturb the fit in
the region and is formed around $\log$ T$_e$ = 5.9 K according to \citet{mazzotta_etal1998}.
We are unable to test 
this hypothesis, however, as there are no suitable atomic data for this line
available in the CHIANTI database. 

The 257.262\,\AA\, line is only just over the 50\% mismatch level so minor blending at the
10--15\% level would be enough to bring the prediction into agreement with the observations.
The NIST database shows \ion{Mg}{8}, \ion{Al}{12}, and \ion{O}{3} lines in the 257.2--257.28\,\AA\,
wavelength range but no suitable atomic data are available to test their contributions.

The worst case for \ion{Fe}{10} is the 195.399\,\AA\, line where the prediction is nearly a factor of 
three too weak suggesting that this line is blended. 
There are a series of \ion{Fe}{10} lines in the region of the line, but including
them in the contribution function and re-computing the DEM leads to a less than 10\% modification
to the predicted intensity. The NIST database shows an \ion{Al}{8} line at 195.374\,\AA\, and
an \ion{Fe}{7} line at 195.391\,\AA\, but no suitable atomic data are available to test 
their contributions. Empirically the \ion{Fe}{7} 195.391\,\AA\, line is found to be a
factor of two stronger than the nearby \ion{Fe}{7} 196.22\,\AA\, line that is unblended
(P.R. Young, private communication). Measuring the 196.22\,\AA\, line 
and correcting by a factor of two leads to a contribution to the 195.399\,\AA\,
intensity of 20--25\%, though the line is very weak in our spectrum. This is not sufficient
to resolve the discrepancy found from the DEM analysis.

In the absence of further information we are unable to determine the sources of these
discrepancies with confidence. The trend for the observed intensities to be larger than 
the predicted ones may suggest that there are additional unknown blends, or that 
the lines are simply 
mis-identified in the spectra because, for example, the wavelengths are assigned to the level transitions
incorrectly. 
There is also always the possibility, however, that the atomic data
are not sufficiently accurate. 

2. The S-like \ion{Fe}{11} 188.299\,\AA\, line is about a factor of two stronger than predicted
by the DEM. It is partially blended with the \ion{Fe}{11} 188.216\,\AA\, line but both
are easily separable when fitted. 
The 188.299\,\AA\, line intensity would need to be blended with
an unknown line of comparable strength to account for the discrepancy but there is no obvious
candidate line to be the blend. The NIST database shows no \ion{Fe}{11}
line at 188.299\,\AA\, and \citet{young_etal2007a} note that the identification of
one or both of these lines is uncertain.
The line at 188.216\,\AA\, is reproduced to high
accuracy by the DEM, however, so this identification looks secure. It could be that the
transition identification or wavelength for the 188.299\,\AA\, line is incorrect in the CHIANTI database.
Alternatively, it may just be a mis-identification.

3. The P-like \ion{Fe}{12} line at 256.925\,\AA\ is significantly
stronger than predicted by the DEM (factor of 25). This large
discrepancy points to significant problems with the atomic data for this line or
that it is mis-identified. 

It has been used on its own here, but the feature also contains 
weaker contributions from two other lines around 256.94\,\AA. Even including these
lines, however, the prediction is still over an order of magnitude out. One possibility
is that there are many more weak contributing lines 
than expected around this wavelength, perhaps from higher level transitions, 
but this is purely conjecture. The current
results suggest that the collisional or radiative rates for this
transition need significant improvement, or that it is mis-identified. 

4. The O-like \ion{Si}{7} 275.352\,\AA\, line is predicted to be a factor of 2.6 higher than observed
and the recently proposed identifications of Ar-like \ion{Fe}{9} lines at 188.485,
189.940, and 197.858\,\AA\, are all predicted to be 65--85\% brighter than observed.
These measurements show a contrasting trend to the cases already discussed i.e. the predicted 
intensities are greater than the 
observed ones, and imply that the presence of unknown blends would only increase the 
discrepancy because the observed intensity would be reduced. Identifying the 189.940\,\AA\,
feature as a line of \ion{Fe}{11} as in \citet{brown_etal2008}, for example, leads to a computed 
intensity that is a
factor of two too high, even without \ion{Fe}{9}. This in turn suggests that there are problems
with the atomic data for these lines, or that the identifications are uncertain. 

As mentioned, there is greater uncertainty in the DEM distribution in the temperature 
range ($\log$ T$_e$ = 5.7--6.1 K) in which these lines are formed. If a local minimum with a 
reduction in the DEM of about a factor of two
is introduced around $\log$ T$_e$ = 5.8 K then the \ion{Fe}{9} lines at
188.485\,\AA\, and 189.940\,\AA\, {\it can} be reproduced at the expense of losing the agreement for 
\ion{Fe}{10} 190.038\,\AA\, and \ion{Fe}{14} 264.787\,\AA. Furthermore, if the minimum is accentuated,
the \ion{Fe}{9} 197.858\,\AA\, line comes into agreement and with a reduction of a factor of four 
even the \ion{Si}{7} 275.352\,\AA\, line can 
be reproduced at the expense
of the \ion{Fe}{8} lines at 185.213\,\AA\, and 186.601\,\AA. 
Therefore, of the eight lines discussed, only
four of any species or ionization state can be reproduced simultaneously. 

We decided to favor the \ion{Fe}{8} lines in this study for five reasons. First, 
the solutions including the  \ion{Fe}{9} and \ion{Si}{7} lines destroy
the agreement found between the current best case EIS DEM distribution and that of our
previous study derived from CDS data because of the minimum around $\log$ T$_e$ = 5.8 K. 
There is no evidence for such a deep local minimum
in the DEM distribution at these temperatures from previous quiet Sun DEM studies, nor
is one expected when one considers the radiative loss curve in this temperature region. 
Second, recent new ionization fraction calculations by \citet{bryans_etal2009} indicate
that the peak fractional abundance for \ion{Fe}{9} is about 20\% too high in the 
\citet{mazzotta_etal1998}
calculations. A corresponding reduction in our predicted intensities would 
bring the \ion{fe}{9} lines to within 30--50\% of the observed intensities and
therefore into closer agreement with the best case EIS DEM emphasizing \ion{Fe}{8}.
Third, the \citet{bryans_etal2009} results also indicate that the \ion{Si}{7} peak
fractional abundance should be increased by a factor of two. This is much larger
than the changes for \ion{Fe}{8} or \ion{Fe}{9} indicating that the \ion{Si}{7}
ionization balance derived by \citet{mazzotta_etal1998} is considerably uncertain.
Fourth, all of the other Si lines used in the study are reproduced satisfactorily without
the minimum. 
Finally, our objective, for this study, is to find a series of reliable Fe lines.

The problems above are also compounded by uncertainty in the formation temperatures of the lines
themselves. 
\citet{young_etal2007b} used a comparison of coronal loop images formed from the \ion{Si}{7} 275.352\,\AA\, line
with images formed from the \ion{Fe}{8} 185.213\,\AA\, line to suggest that \ion{Fe}{8}
was actually formed at a higher temperature ($\log$ T$_e$ = 5.8 K) than the 
\citet{mazzotta_etal1998} ionization balance calculations would suggest ($\log$ T$_e$ = 5.6 K).
Their argument was based on the fact that the atomic structure of Fe is more complex than
that of Si, but this issue requires further attention because of the potential problem found here
with \ion{Si}{7} 275.352\,\AA. 
\citet{brown_etal2008} identify seven \ion{Fe}{8} lines and five \ion{Si}{7} lines
in their EIS quiet Sun spectrum. 
If the temperature of formation of these lines in ionization equilibrium
is the problem they should all be shifted in the same way.
An inspection of images in all these lines, or a more quantitative comparison of, e.g.,
relative intensity profiles with location \citep{brooks_etal1999}, may firm up conclusions
on this issue. 

\subsection{DEM Subset and Iron Only}
\label{demsubset}
We have applied the best case EIS DEM solution for the lines in Table \ref{tab1}
to the complete dataset of 77 fitted spectral lines
to obtain a global view of the reliability of the other lines, and to confirm the identification of
single reliable lines 
at each temperature that could be used to derive a DEM curve with minimal measurements. Our 
expectation, of course, was that many of these lines would come from the set already used for our 
best case solution except for cases where another line was needed to fill in a temperature interval. 

Figure \ref{fig5} shows the result of that application in the same format as 
Figure \ref{fig3}. 
Empty colored circles now indicate lines that were fitted and reproduced by the DEM, but were deselected
because a better candidate line existed at the same temperature. Before referring to that discussion, however, 
note the significant spread in observed to predicted ratios for many of the lines. In many cases, the
predicted intensities are factors of two or more out. 
A possible explanation for these discrepancies is that
the lines are weak or blended with unidentified lines. A more detailed analysis of the weaker lines
would be needed to confirm or reject this suggestion.

Having commented on that possibility, however, there are many cases where the error in the Gaussian fit is
relatively large compared to that of the lines in the initial selection (Table \ref{tab2}), but the fit itself still 
looks very good. 
We believe, therefore, that many additional weaker lines in the EIS spectra could be used usefully for future studies.
The available atomic data for many lines, however, are not of sufficient accuracy at this stage for that purpose
(especially lines of Fe).
As pointed out earlier and by others \citep{lanzafame_etal2002,young_etal2007b} the atomic modeling for complex
species such as Fe needs significant improvement, especially for the analysis of weaker low signal-to-noise
lines presumably emitted from transitions between higher levels where the population structure needs more
detailed modeling. State-of-the-art 
solar spectroscopic observations continue to provide useful guidance for the direction of modeling efforts. 

The filled circles in Figure \ref{fig5} now represent the best reliable lines selected at each temperature.
In making that choice, stronger lines were preferred over weaker lines (if two lines were reproduced to the
same accuracy level) and Fe lines were preferred over other species (in an attempt to provide a subset
of Fe lines that could be used to obtain the DEM free of uncertainties in the elemental abundances).

The final subset of selected lines is given in Table \ref{tab2} and they are mostly strong Fe lines.
One Mg and one Al line are also included to fill out the temperature range. 

The subset of lines clearly works. It spans the temperature interval $\log$ T$_e$ = 5.55--6.35 K
and can reproduce the best case EIS DEM of Figure \ref{fig4} well with a comparable reduction in the abundance
of Fe to that adopted in \S \ref{resbcd}. The Al and Mg lines can also be brought into agreement
with the others by decreasing their elemental abundances by 20--30\%. Since 
these abundance adjustments are based only on the measurement of one line we consider this solution unsatisfactory.
In fact, if we decrease the values of the magnitude of the DEM at each temperature by 40\%
then all the lines can be reproduced 
to within 20\% without the need for adjusting any elemental abundances. 
This is within the stated accuracy of the EIS photometric calibration measured in
the laboratory pre-launch \citep{lang_etal2006}. 

Since many of the selected lines reproduced accurately in this study and shown
in Tables \ref{tab1} and \ref{tab2} span both the short and long wavelengths of both EIS CCDs A and B, the accuracy
of their reproduction essentially provides a preliminary validation of the EIS photometric calibration
on orbit. More work and detailed analysis across the complete wavelength range is needed, however.
 
\subsection{Single Pixel DEM}
\label{demsinglepix}
In the previous section we derived the average DEM distribution for this region using a subset of 
Fe, Mg, and Al lines (see Table \ref{tab2}). Here we use only the subset of Fe lines 
to demonstrate the sensitivity of EIS by deriving the same DEM distribution from
a single EIS pixel. To do so, we applied the average DEM solution to the intensity measurements from
the single pixel (marked with small boxes on the images in Figure \ref{fig2}). The Fe abundance was also increased
by 10\%.

Figure \ref{fig6} shows the resulting DEM distribution as a function of temperature. The square boxes
show the positions of the spline knots used to represent the DEM. The
red dots show the ratios of the observed to predicted intensities scaled by the value of the DEM at the
formation temperature of the line. As before, the closer these lines are to the
DEM curve the better they are reproduced. 
The quantitative values from the analysis are given in Table \ref{tab3}, which is in the same
format as Tables \ref{tab1} and \ref{tab2}. It can be seen that each of the dots lies nicely on the curve. 
The uncertainties in the fits to the line intensities are generally increased compared to the spectrum
averaged over the full FOV, being factors of approximately 2--3 larger, though they are still close to our tolerance
of about 30\% except
for the low signal-to-noise \ion{Fe}{15} 284.160\,\AA\, line. The uncertainty in the fitted 
intensity of the latter line becomes comparable to that for \ion{Fe}{14} 274.203\,\AA\, if    
we average the spectrum over 2--4 pixels (depending on the quality of the spectra).

The Fe lines span the temperature interval $\log$ T$_e$ = 5.55--6.35 K and they are
all reproduced to within 16\%. This result demonstrates the aptitude of EIS for studies of 
inhomogeneities in the DEM distribution at high spatial resolution. 

\subsection{Single Pixel DEM with MCMC algorithm}
\label{demsinglepixelmcmc}
As an independent verification of the results we also applied a Monte Carlo Markov Chain (MCMC)
emission measure algorithm \citep{kashyap&drake_1998} to the data. The code is available
in the PINTofALE IDL analysis package \citep{kashyap&drake_2000}. The derived DEM distribution
is shown in Figure \ref{fig7} (histogram plot), together with the DEM loci curves
derived from the results for each spectral line using the formulation 
\begin{equation}
EM (T_e) = I_{n l} / ( G(T_e,N_0) \times T_e )
\end{equation} 
Here, $N_0$ is the measured value of electron density for this dataset. The MCMC DEM results 
were used to predict the spectral line intensities and the
results are given in the final two
columns of Table \ref{tab3}. These are very close to the results obtained using the
spline method, except for \ion{Fe}{14} 274.203\,\AA\, which is about 60\% too low. This difference
could indicate that the emission measure should be increased at the formation temperature of
\ion{Fe}{14}, which could be evidence of a high temperature tail in the DEM distribution. This
possibility is discussed in detail in \citet{warren&brooks_2009}. 

\section{Uniformity of the quiet Sun DEM distribution}
An immediate advantage of obtaining a reliable subset of Fe lines is that we can
assess the generality of our results quickly. 
We compared the averaged intensities for the Fe lines recommended in the final subset 
over a 20 hour period surrounding the main observations used in this paper on 2007 January 30--31. All
the data were taken using the same EIS study as analyzed
here. The observations were also taken with a fixed pointing, so the observed area shifted significantly
during this time. 
We found that the intensities of these Fe lines varied by less than 18\% 
between the datasets and that they can therefore be reproduced with the same DEM curve as derived
for the main dataset. 

To explore the uniformity of the quiet Sun DEM distribution in general, we collected 39 additional datasets
taken between 2007 January 25 and April 14. These latter datasets were obtained with a different 
observing sequence with study acronym SYNOP001. 
This study takes two 40$''$ slot images followed 
by two 1$''$ slit exposures with times of 30s and 90s, respectively. It also returns full
CCD spectra, but the FOV is only 1$'' \times$ 256$''$ (for the slit exposures). Here we used the 90s 
exposure from each dataset and averaged the spectra over the full 256$''$ slit direction. 

SYNOP001 is nominally run on quiet Sun near disk
center to monitor the EIS sensitivity. In the early stages of the mission, however, it was also used
to obtain full CCD spectra of other targets. By checking EIT and 
{\it Hinode} X-ray Telescope \citep[XRT,][]{golub_etal2007} full Sun images from 2007 January
to April, we were able to select time periods when solar disk center was free of any influence
from active regions or equatorial coronal holes. The 39 datasets represent the result of that
verification after also ignoring 
observations that were pointed far from disk center. We then reduced the datasets as before
and fitted the subset of Fe lines in order to perform a DEM analysis on all
45 datasets. The results are shown in Table \ref{tab4}.

Table \ref{tab4} gives the date and start time of the observations in columns 2 and 3. The HPW001\_FULLCCD\_RAST
datasets are numbered 10--15 in column 1. All the other datasets are from SYNOP001 observations. 
Our goal
was to check if the shape of the DEM distribution changed significantly from quiet Sun region to region. So we 
initially applied our best-case solution to the averaged spectrum of each dataset to see whether the observed line intensities 
could be reproduced. If not, we then adjusted the DEM curve. Since the shape of the DEM is the key point
of interest here, we scaled the DEM by an adjustable constant factor to try to match the intensities. We
only modified the shape if it became necessary. This constant factor ($\alpha$) is given in column 4
of the table and varies by up to a factor of 2 in the regions we have analyzed. This factor is a measure
of the relative difference between the DEM curves on an absolute magnitude scale and may be representative
of the amount of heating in each region. Alternatively, it could indicate a deviation from the adopted 
Fe abundance (in fact $\alpha$ was derived by adjusting this abundance). A mean dataset created by
averaging all the intensities in each column (i.e. averaging over all the datasets) is also added in the
last row of the table. 

Columns 6--12 give the observed intensities of the 7 Fe lines from the subset, and in brackets we 
show the ratios of predicted to observed intensities from the DEM solutions we derived. The measurement
uncertainties from the spectral fits are not shown to preserve table space. Of the 315 intensities
shown in the table only 3 are reproduced to worse than 30\%. All 3 cases are \ion{Fe}{14} 274.203\,\AA\,
intensities and in all 3 cases the predicted intensity is too low. These cases behave like the MCMC
solution presented in \S \ref{demsinglepixelmcmc} and may be further evidence of a high temperature tail
in the distribution. It is also possible, however, to modify the high temperature distribution    
to reproduce this intensity at the expense of other lines.

Figure \ref{fig8} shows a plot of all the DEM solutions derived from all of the datasets listed in Table \ref{tab4}.
This figure clearly illustrates the similarity in the shapes of the curves.
In \S \ref{resbcd} we showed that the \citet{brooks&warren_2006}
and EIS DEMs maintained similar shapes up to $\log$ $T_e$ = 6.2 K when the lines become weak and
the curves start to fall rapidly. Figure \ref{fig8} demonstrates that this is true in general. All the
curves maintain similar shapes up to at least $\log$ $T_e$ = 6.1. 

We calculated ratios of each DEM solution to the distribution derived from the
January 30 11:19UT dataset as a function of temperature and
measured the temperature at which
the ratio for each curve deviated by more than 50\%. These values are given in column 5 of Table \ref{tab4}. The vast majority
(93\%) of the curves do not deviate by more than 50\% until $\log$ $T_e$ = 6.15 K, and two thirds of the curves are within this limit
until $\log$ $T_e$ =  6.2 K. Note that in many cases the January 30 distribution is able 
to reproduce all of the intensities. The eye is naturally drawn to the spread in the curves at high temperature in Figure \ref{fig8}, but a value of 6.8 in
the Table indicates that the curve never deviates by more than 50\%. 

\section{Summary and Discussion}
\label{conc}
We have performed a DEM analysis of a quiet coronal region using observations obtained by {\it Hinode}/EIS.
We find that the well-known quiet coronal DEM distribution can be recovered as expected, and that the intensities
of the majority of the reliable lines, averaged over the full FOV, can be reproduced to within 30\%.
Only about one quarter of the spectral line intensities are reproduced to worse accuracy than 50\%,
and this result is comparable to that achieved for previous EUV spectrometers, for example, CDS
\citep{brooks&warren_2006}.

The analysis also provides clues to the explanations for the discrepancies found for some lines that
could be exploited in the future. Despite recent improvements in the fundamental atomic data available
for \ion{Fe}{12}, for example, problems remain for some lines (e.g.,
\ion{Fe}{12} 256.925\,\AA). As suggested by previous authors (\citealt{lanzafame_etal2002,young_etal2009}), further improvement
in the atomic modeling of complex species such as Fe is necessary to fully exploit the high quality
EIS observations.

We present also a subset of nine EIS lines spanning the temperature interval $\log$ T$_e$ = 5.55--6.35 K
that can be used to derive the DEM distribution reliably. This includes a series of seven Iron lines
that can be used to derive the DEM distribution free from the possibility of uncertainties in the
elemental abundances. This subset is likely to be useful for extensive studies of, e.g., 
the temporal evolution of the DEM during active events.

We have also demonstrated the power of EIS for studies of spatial inhomogeneities in
the quiet coronal DEM distribution by deriving the DEM curve from a single EIS
pixel. The curve was shown to follow the same shape as the DEM curve 
derived from a spectrum produced by averaging the whole observed field-of-view.
All of the observed intensities from a subset of Fe lines are reproduced to within
16\%. This is in fact better than the result obtained for the averaged spectrum, though
the fitted intensities have larger uncertainties.

Since many of the lines 
in Tables \ref{tab1} and \ref{tab2} are reproduced to within the accuracy of the pre-launch laboratory calibration
of EIS and 
span both the short and long wavelength ranges of both CCDs, this
analysis also provides a preliminary validation of the photometric calibration of EIS on orbit,
that was realized to unprecedented accuracy in the laboratory. 
Thus the pre-launch calibration
can be used with confidence. Regular observations are carried out by the EIS team to monitor any
degradation in sensitivity with time.

The EIS DEM is very similar to that derived in \citet{brooks&warren_2006} using CDS data. 
They were derived independently from data
taken nearly 9 years apart,
using completely different spectral lines, in completely different EUV wavelength ranges, with their own associated
atomic data uncertainties and instrument specific issues, yet 
the two curves maintain basically the same shape in the temperature interval $\log$ T$_e$ = 5.4--6.2 K.
We confirmed that this result was true in general by analyzing 45 EIS quiet Sun datasets taken
in the period 2007 January to April. 

Attempts to use the DEM to infer information about the heating have a long history,
see e.g. \citet{jordan_1980}, and it is well known that
hydrostatic simulations of uniformly heated coronal loops fail to correctly model
the DEM distribution \citep[][and references therein]{schrijver&aschwanden_2002}. In
particular, the slope of the DEM below the peak is much steeper in the observations. 
Using loops with expanding cross-section, or incorporating loop foot-point heating, 
improves the agreement but does not solve the problem. Hydrodynamic modeling may
provide further insight, but introduces flows that are assumed to be negligible in
the DEM method. Nevertheless, 
the observation of a {\it universal} DEM throughout the quiet Sun (even down
to small spatial scales) is 
potentially an important result because it provides a further constraint on the heating mechanism.

The shape of the DEM as a function of 
temperature is dependent on several terms in the energy balance equation, and 
the relative importance of
flows, loop geometry, the heating mechanism, and cooling by radiation and
conduction is not known. 
One possibility is that although the absolute magnitude of
the quiet Sun DEM may scale with the amount of energy released in each region, 
the {\it shape} of the distribution 
is mainly a function of the radiating and conducting properties of the plasma, and
is fairly insensitive to the location and rate of energy deposition.
Since these functions have a fixed shape, the DEM would also have a fixed
shape if they dominate the energy balance. 
The loop geometry and the heating mechanism could, in principle,
vary from location to location, but if they are the dominant factors controlling the DEM shape then the observation
that the DEM does not vary would imply that these factors also do not vary substantially from region to region. 
We expect that there are several heating mechanisms operating in the quiet Sun, so these observations imply
either that one is dominant, or that their influence on the DEM shape is negligible.

\acknowledgments
We thank John Mariska and Peter Young for helpful comments and suggestions.
We also thank the referee for constructive suggestions on how to physically
interpret the results.
DHB and HPW acknowledge funding support
from the NASA Hinode program. 
CHIANTI is a collaborative project involving the NRL (USA), RAL (UK), and the following Universities:
College London (UK), Cambridge (UK), George Mason (USA), and Florence (Italy).
{\it Hinode} is a Japanese mission developed and launched by ISAS/JAXA,
with NAOJ as domestic partner and NASA and STFC (UK) as international partners.
It is operated by these agencies in co-operation with ESA and NSC (Norway).

{\it Facilities:} \facility{Hinode (EIS), SOHO (EIT) }     

\bibliography{/home/brooks/latex/dhb_bib/solar}
\begin{deluxetable}{lccr@{$\pm$}lcc}
\tabletypesize{\footnotesize}
\tablewidth{0pt}
\tablecaption{Quiet Sun intensities from the EIS Spectrometer.}
\tablehead{
\multicolumn{1}{l}{Ion} &
\multicolumn{1}{c}{$\lambda_{obs}$ (\AA)} &
\multicolumn{1}{c}{$\log T_{max}$ (K)} &
\multicolumn{2}{c}{I$_{obs}^a$} &
\multicolumn{1}{c}{I$_{dem}^a$} &
\multicolumn{1}{c}{Ratio}
}
\tablenotetext{\textit{a}}{Units are erg cm$^{-2}$ s$^{-1}$ sr$^{-1}$.}
\startdata
  \ion{FE}{8} & 185.213 & 5.55 &    19.73 & 0.06 &   20.81 & 1.05 \\
  \ion{FE}{8} & 186.601 & 5.55 &    14.77 & 0.05 &   14.60 & 0.99 \\
  \ion{Si}{7} & 275.352 & 5.70 &    10.52 & 0.04 &   27.08 & 2.57 \\
  \ion{FE}{9} & 188.485 & 5.80 &    14.13 & 0.03 &   23.27 & 1.65 \\
  \ion{FE}{9} & 189.940 & 5.80 &     8.24 & 0.05 &   13.42 & 1.63 \\
  \ion{FE}{9} & 197.858 & 5.80 &     8.78 & 0.03 &   16.06 & 1.83 \\
 \ion{FE}{10} & 177.239 & 6.00 &   109.74 & 0.52 &  140.64 & 1.28 \\
 \ion{FE}{10} & 184.537 & 6.00 &    57.21 & 0.09 &   55.72 & 0.97 \\
 \ion{FE}{10} & 190.038 & 6.00 &    20.34 & 0.07 &   15.67 & 0.77 \\
 \ion{FE}{10} & 193.715 & 6.00 &     4.72 & 0.02 &    2.79 & 0.59 \\
 \ion{FE}{10} & 195.399 & 6.00 &     4.35 & 0.03 &    1.48 & 0.34 \\
 \ion{FE}{10} & 257.262 & 6.00 &    47.66 & 0.08 &   31.11 & 0.65 \\
 \ion{FE}{11} & 180.401 & 6.05 &   190.56 & 0.30 &  185.56 & 0.97 \\
 \ion{FE}{11} & 182.167 & 6.05 &    21.95 & 0.11 &   24.85 & 1.13 \\
 \ion{FE}{11} & 188.216 & 6.05 &    93.95 & 0.10 &   87.57 & 0.93 \\
 \ion{FE}{11} & 188.299 & 6.05 &    64.03 & 0.10 &   31.88 & 0.50 \\
  \ion{S}{10} & 264.233 & 6.10 &    14.26 & 0.05 &   11.35 & 0.80 \\
 \ion{FE}{12} & 192.394 & 6.15 &    36.96 & 0.04 &   48.48 & 1.31 \\
 \ion{FE}{12} & 193.509 & 6.15 &    79.57 & 0.06 &  102.19 & 1.28 \\
 \ion{FE}{12} & 195.119 & 6.15 &   135.38 & 0.05 &  151.45 & 1.12 \\
 \ion{FE}{12} & 256.925 & 6.15 &    13.62 & 0.06 &    0.68 & 0.05 \\
 \ion{Si}{10} & 258.375 & 6.15 &    31.84 & 0.07 &   28.12 & 0.88 \\
 \ion{Si}{10} & 261.058 & 6.15 &    16.61 & 0.05 &   16.49 & 0.99 \\
 \ion{Si}{10} & 271.990 & 6.15 &    15.57 & 0.05 &   12.39 & 0.80 \\
 \ion{Si}{10} & 277.255 & 6.15 &    10.06 & 0.05 &   10.14 & 1.01 \\
 \ion{FE}{13} & 202.044 & 6.20 &    82.05 & 0.09 &   80.63 & 0.98 \\
 \ion{FE}{14} & 264.787 & 6.25 &     9.43 & 0.05 &    7.23 & 0.77 \\
 \ion{FE}{14} & 274.203 & 6.25 &    11.26 & 0.05 &   10.76 & 0.96 \\
 \ion{FE}{15} & 284.160 & 6.35 &    13.16 & 0.08 &   15.18 & 1.15 \\
\enddata
\label{tab1}
\end{deluxetable}
\begin{deluxetable}{lccr@{$\pm$}lcc}
\tabletypesize{\footnotesize}
\tablewidth{0pt}
\tablecaption{Quiet Sun intensities from the EIS Spectrometer.}
\tablehead{
\multicolumn{1}{l}{Ion} &
\multicolumn{1}{c}{$\lambda_{obs}$ (\AA)} &
\multicolumn{1}{c}{$\log T_{max}$ (K)} &
\multicolumn{2}{c}{I$_{obs}^a$} &
\multicolumn{1}{c}{I$_{dem}^a$} &
\multicolumn{1}{c}{Ratio}
}
\tablenotetext{\textit{a}}{Units are erg cm$^{-2}$ s$^{-1}$ sr$^{-1}$.}
\startdata
  \ion{FE}{8} & 185.213 & 5.55 &    19.73 & 0.06 &   18.44 & 0.93 \\
  \ion{Mg}{6} & 268.986 & 5.60 &     2.72 & 0.08 &    2.50 & 0.92 \\
 \ion{FE}{10} & 184.537 & 6.00 &    57.21 & 0.09 &   49.38 & 0.86 \\
 \ion{FE}{11} & 188.216 & 6.05 &    93.95 & 0.10 &   77.61 & 0.83 \\
  \ion{Al}{9} & 284.015 & 6.05 &     2.89 & 0.09 &    3.36 & 1.16 \\
 \ion{FE}{12} & 195.119 & 6.15 &   135.38 & 0.05 &  134.23 & 0.99 \\
 \ion{FE}{13} & 202.044 & 6.20 &    82.05 & 0.09 &   71.47 & 0.87 \\
 \ion{FE}{14} & 274.203 & 6.25 &    11.26 & 0.05 &    9.54 & 0.85 \\
 \ion{FE}{15} & 284.160 & 6.35 &    13.16 & 0.08 &   13.46 & 1.02 \\
\enddata
\label{tab2}
\end{deluxetable}
\begin{deluxetable}{lccr@{$\pm$}lcccc}
\tabletypesize{\footnotesize}
\tablewidth{0pt}
\tablecaption{Quiet Sun intensities from a single EIS pixel.}
\tablehead{
\multicolumn{1}{l}{Ion} &
\multicolumn{1}{c}{$\lambda_{obs}$ (\AA)} &
\multicolumn{1}{c}{$\log T_{max}$ (K)} &
\multicolumn{2}{c}{I$_{obs}^a$} &
\multicolumn{1}{c}{I$_{dem}^a$} &
\multicolumn{1}{c}{Ratio} &
\multicolumn{1}{c}{I$_{dem}^b$} &
\multicolumn{1}{c}{Ratio$^b$}
}
\tablenotetext{\textit{a}}{Units are erg cm$^{-2}$ s$^{-1}$ sr$^{-1}$.}
\tablenotetext{\textit{b}}{Results using MCMC algorithm.}
\startdata
  \ion{Fe}{8} & 185.213 & 5.55 &    18.63 & 5.40 &   20.26 & 1.09  &          18.98  &   1.02 \\
 \ion{Fe}{10} & 184.536 & 6.00 &    58.01 &    11.02 &   54.27 & 0.94  &      65.05  &   1.12 \\
 \ion{Fe}{11} & 188.216 & 6.05 &    99.04 & 9.90 &   85.29 & 0.86  &       92.23  &   0.93 \\
 \ion{Fe}{12} & 195.119 & 6.15 &   129.65 & 6.48 &  147.50 & 1.14  &       128.06 &    0.99 \\
 \ion{Fe}{13} & 202.044 & 6.20 &    80.11 &    10.41 &   78.53 & 0.98  &      88.40  &   1.10 \\
 \ion{Fe}{14} & 274.203 & 6.25 &    12.03 & 4.09 &   10.48 & 0.87  &       7.37   &  0.61 \\
 \ion{Fe}{15} & 284.160 & 6.35 &    13.65 & 7.64 &   14.79 & 1.08  &       15.33  &   1.12  \\
\enddata
\label{tab3}
\end{deluxetable}
\begin{deluxetable}{lcccrrrrrrrr}
\rotate        
\tabletypesize{\scriptsize}
\tablewidth{0pt}
\tablecaption{Quiet Sun intensities from 45 datasets.}
\tablehead{
\multicolumn{1}{l}{} &
\multicolumn{4}{c}{} &
\multicolumn{6}{c}{I$_{obs}^a$ (I$_{dem}$/I$_{obs}$)} \\
[.3ex]\cline{6-12}\\[-1.6ex]
\multicolumn{1}{l}{}  &
\multicolumn{4}{c}{}  &
\multicolumn{1}{c}{\ion{Fe}{8}}  &
\multicolumn{1}{c}{\ion{Fe}{10}}  &
\multicolumn{1}{c}{\ion{Fe}{11}}  &
\multicolumn{1}{c}{\ion{Fe}{12}} &
\multicolumn{1}{c}{\ion{Fe}{13}} &
\multicolumn{1}{c}{\ion{Fe}{14}} &
\multicolumn{1}{c}{\ion{Fe}{15}} \\
[.3ex]\cline{6-12}\\[-1.6ex]
\multicolumn{1}{l}{No.} &
\multicolumn{1}{c}{Date} &
\multicolumn{1}{c}{Time$^b$} &
\multicolumn{1}{c}{$\alpha^d$} &
\multicolumn{1}{c}{T$_{cut}^e$} &
\multicolumn{1}{c}{185.213$\lambda$} &
\multicolumn{1}{c}{184.536$\lambda$} &
\multicolumn{1}{c}{188.216$\lambda$} &
\multicolumn{1}{c}{195.119$\lambda$} &
\multicolumn{1}{c}{202.044$\lambda$} &
\multicolumn{1}{c}{274.203$\lambda$} &
\multicolumn{1}{c}{284.160$\lambda$}
}
\tablenotetext{\textit{a}}{Units are erg cm$^{-2}$ s$^{-1}$ sr$^{-1}$.}
\tablenotetext{\textit{b}}{Start time of the first exposure in the raster.}
\tablenotetext{\textit{c}}{Units are arcsec.}
\tablenotetext{\textit{d}}{Scaling factor for the Fe abundance (see text).}
\tablenotetext{\textit{e}}{Temperature at which the DEM deviates from the best-case by more than 50\%.}
\startdata
 1 & 25-Jan-07 & 17:51:36 & 1.5 & 6.80 &  21.01 (1.28) & 58.91 (1.23) & 96.19 (1.18) &193.38 (1.01) &131.29 (0.80) & 17.78 (0.78) & 23.43 (0.84)\\
 2 & 26-Jan-07 & 00:06:50 & 1.0 & 6.80 &  17.58 (1.05) & 48.36 (1.02) & 74.22 (1.05) &115.91 (1.16) & 66.73 (1.07) & 11.39 (0.84) & 14.25 (0.94)\\
 3 & 26-Jan-07 & 06:04:49 & 1.1 & 6.25 &  15.37 (1.27) & 44.67 (1.22) & 84.97 (1.03) &140.94 (1.11) & 79.65 (1.11) & 16.10 (0.79) & 19.21 (1.05)\\
 4 & 26-Jan-07 & 12:09:28 & 0.8 & 6.80 &  18.24 (0.79) & 43.65 (0.89) & 70.09 (0.87) &104.28 (1.01) & 62.76 (0.89) &  8.42 (0.89) &  8.29 (1.27)\\
 5 & 26-Jan-07 & 18:09:50 & 0.8 & 6.80 &  19.82 (0.77) & 49.92 (0.82) & 76.99 (0.84) &110.17 (1.01) & 56.40 (1.05) &  8.56 (0.93) &  8.60 (1.30)\\
 6 & 27-Jan-07 & 06:28:49 & 0.9 & 6.40 &  21.60 (0.77) & 53.65 (0.83) & 90.47 (0.77) &107.10 (1.11) & 50.77 (1.22) &  8.06 (1.00) &  8.35 (1.30)\\
 7 & 27-Jan-07 & 11:59:28 & 0.7 & 6.30 &  17.61 (0.78) & 43.97 (0.84) & 74.47 (0.78) &101.45 (0.97) & 52.02 (0.97) &  7.57 (0.84) &  6.47 (1.27)\\
 8 & 28-Jan-07 & 00:02:50 & 1.2 & 6.10 &  29.91 (0.81) & 74.39 (0.77) & 80.11 (0.97) & 85.14 (1.29) & 33.67 (1.30) &  5.65 (0.68) &  3.38 (0.93)\\
 9 & 28-Jan-07 & 06:02:49 & 1.1 & 6.10 &  23.54 (0.88) & 63.13 (0.77) & 71.74 (0.85) & 62.62 (1.23) & 20.20 (1.31) &  2.99 (0.59) &  1.16 (0.77)\\
10 & 30-Jan-07 & 11:19:12 & 1.1 & 6.80 &  19.15 (1.04) & 54.69 (0.98) & 94.47 (0.89) &140.57 (1.03) & 84.21 (0.92) & 11.15 (0.93) & 12.74 (1.14)\\
11 & 30-Jan-07 & 14:34:40 & 1.1 & 6.80 &  19.85 (1.03) & 57.51 (0.95) & 94.86 (0.91) &138.27 (1.08) & 82.98 (0.95) & 11.27 (0.94) & 13.63 (1.09)\\
12 & 30-Jan-07 & 17:50:08 & 1.0 & 6.80 &  19.66 (0.94) & 54.67 (0.91) & 85.31 (0.91) &122.89 (1.09) & 74.59 (0.96) & 10.50 (0.91) & 13.90 (0.97)\\
13 & 30-Jan-07 & 21:05:35 & 1.0 & 6.80 &  18.52 (1.00) & 52.60 (0.94) & 83.26 (0.93) &129.61 (1.04) & 78.02 (0.92) &  9.67 (0.99) & 12.54 (1.07)\\
14 & 31-Jan-07 & 00:21:02 & 1.1 & 6.80 &  18.83 (1.11) & 56.82 (0.98) & 97.46 (0.90) &145.76 (1.04) & 87.77 (0.92) & 11.66 (0.92) & 15.26 (1.00)\\
15 & 31-Jan-07 & 03:36:29 & 1.2 & 6.80 &  19.05 (1.16) & 62.54 (0.95) &108.96 (0.85) &162.39 (0.99) & 94.58 (0.91) & 11.73 (0.97) & 14.55 (1.11)\\
16 & 31-Jan-07 & 12:01:59 & 1.5 & 6.80 &  25.63 (1.08) & 80.04 (0.92) &123.75 (0.94) &188.82 (1.06) &110.54 (0.97) & 15.07 (0.95) & 19.89 (1.01)\\
17 & 15-Mar-07 & 11:19:49 & 1.2 & 6.15 &  20.47 (1.11) & 68.23 (0.84) &103.42 (0.79) &131.85 (0.93) & 57.47 (0.88) &  5.41 (0.82) &  2.67 (1.28)\\
18 & 15-Mar-07 & 23:57:28 & 1.4 & 6.15 &  21.45 (1.23) & 77.09 (0.86) &119.62 (0.80) &163.32 (0.89) & 75.03 (0.82) &  7.22 (0.77) &  3.44 (1.30)\\
19 & 16-Mar-07 & 11:09:28 & 1.4 & 6.15 &  21.89 (1.21) & 77.33 (0.88) &120.65 (0.82) &174.45 (0.87) & 80.13 (0.80) &  7.82 (0.74) &  3.59 (1.29)\\
20 & 16-Mar-07 & 18:01:27 & 1.5 & 6.15 &  25.29 (1.13) & 79.02 (0.90) &102.96 (0.97) &126.88 (1.14) & 57.30 (1.00) &  6.31 (0.72) &  2.32 (1.28)\\
21 & 17-Mar-07 & 06:07:27 & 1.8 & 6.30 &  26.30 (1.26) & 96.63 (0.91) &168.65 (0.81) &250.97 (0.92) &122.29 (0.97) & 17.13 (0.87) & 15.22 (1.28)\\
22 & 17-Mar-07 & 09:51:43 & 1.3 & 6.35 &  14.78 (1.28) & 62.52 (1.05) &137.71 (0.81) &235.86 (0.84) &136.43 (0.77) & 16.67 (0.82) & 14.28 (1.25)\\
23 & 17-Mar-07 & 18:03:47 & 1.5 & 6.80 &  17.64 (1.29) & 63.94 (1.12) &122.23 (0.94) &209.65 (0.95) &130.62 (0.81) & 17.77 (0.80) & 18.96 (1.05)\\
24 & 18-Mar-07 & 10:05:53 & 1.0 & 6.20 &  17.85 (1.04) & 56.69 (0.86) & 89.93 (0.82) &132.11 (0.91) & 73.71 (0.79) &  8.14 (0.82) &  5.94 (1.28)\\
25 & 19-Mar-07 & 11:20:29 & 1.8 & 6.20 &  20.01 (1.26) & 68.41 (1.27) &130.33 (1.16) &254.94 (1.16) &211.88 (0.90) & 42.41 (0.79) & 78.82 (0.91)\\
26 & 19-Mar-07 & 18:06:29 & 1.2 & 6.15 &  16.67 (1.28) & 52.31 (1.19) &101.49 (1.08) &231.45 (0.99) &217.55 (0.77) & 43.39 (0.80) & 91.73 (1.04)\\
27 & 20-Mar-07 & 00:12:51 & 1.4 & 6.15 &  19.20 (1.30) & 65.59 (1.14) &116.38 (1.13) &243.21 (1.10) &216.55 (0.83) & 42.65 (0.77) & 84.74 (0.86)\\
28 & 20-Mar-07 & 06:13:28 & 1.4 & 6.15 &  19.53 (1.28) & 70.57 (1.02) &136.74 (0.91) &264.89 (0.96) &215.37 (0.82) & 43.86 (0.78) & 94.06 (0.90)\\
29 & 22-Mar-07 & 12:55:59 & 1.5 & 6.30 &  22.99 (1.17) & 71.71 (1.00) &114.29 (0.98) &164.63 (1.18) & 81.25 (1.30) & 17.90 (0.83) & 29.64 (0.77)\\
30 & 23-Mar-07 & 00:04:50 & 1.0 & 6.20 &  22.97 (0.81) & 53.23 (0.91) & 79.47 (0.92) &111.87 (1.07) & 59.19 (0.97) &  8.10 (0.80) &  5.62 (1.29)\\
31 & 23-Mar-07 & 06:17:58 & 1.2 & 6.20 &  23.50 (0.94) & 66.33 (0.87) &113.32 (0.78) &145.81 (0.99) & 65.17 (1.09) & 10.17 (0.82) &  7.56 (1.28)\\
32 & 23-Mar-07 & 06:26:50 & 1.2 & 6.25 &  18.39 (1.21) & 61.21 (0.96) &108.83 (0.83) &167.49 (0.90) & 93.18 (0.82) & 12.24 (0.77) &  9.04 (1.30)\\
33 & 23-Mar-07 & 11:55:20 & 1.1 & 6.80 &  16.95 (1.25) & 54.17 (1.04) & 97.35 (0.91) &157.30 (0.98) & 95.38 (0.86) & 13.38 (0.82) & 13.18 (1.17)\\
34 & 23-Mar-07 & 18:15:20 & 1.6 & 6.15 &  21.03 (1.28) & 86.95 (0.89) &150.18 (0.78) &211.77 (0.88) & 98.73 (0.87) & 11.13 (0.81) &  7.03 (1.29)\\
35 & 25-Mar-07 & 00:04:05 & 1.5 & 6.80 &  21.15 (1.30) & 82.28 (0.90) &135.97 (0.86) &193.44 (1.04) &103.56 (1.04) & 17.75 (0.81) & 26.44 (0.77)\\
36 & 25-Mar-07 & 10:45:49 & 1.3 & 6.15 &  25.40 (0.90) & 82.74 (0.78) &119.75 (0.80) &151.10 (0.98) & 65.12 (1.01) &  8.09 (0.82) &  5.01 (1.30)\\
37 & 25-Mar-07 & 18:06:19 & 1.3 & 6.20 &  19.13 (1.26) & 78.39 (0.80) &114.16 (0.83) &138.90 (1.10) & 65.59 (1.10) &  7.52 (1.07) &  6.88 (1.28)\\
38 & 04-Apr-07 & 05:37:19 & 1.8 & 6.80 &  25.85 (1.28) & 85.28 (1.04) &150.05 (0.93) &245.04 (0.98) &148.01 (0.87) & 19.22 (0.89) & 21.91 (1.10)\\
39 & 09-Apr-07 & 23:57:08 & 1.5 & 6.80 &  20.69 (1.30) & 64.73 (1.12) &110.00 (1.04) &194.13 (1.03) &135.27 (0.79) & 18.83 (0.77) & 19.99 (1.06)\\
40 & 10-Apr-07 & 06:12:49 & 1.3 & 6.15 &  15.14 (1.29) & 64.70 (1.06) &142.88 (0.83) &254.14 (0.89) &168.95 (0.79) & 25.65 (0.77) & 32.57 (0.95)\\
41 & 11-Apr-07 & 13:23:38 & 1.4 & 6.15 &  26.37 (1.01) & 77.26 (0.84) &100.50 (0.89) &105.97 (1.19) & 37.56 (1.28) &  4.17 (0.87) &  1.84 (1.19)\\
42 & 12-Apr-07 & 07:30:30 & 2.0 & 6.10 &  43.51 (0.92) &108.17 (0.78) &118.91 (0.89) &122.53 (1.12) & 41.64 (1.30) &  9.21 (0.57) &  5.34 (1.02)\\
43 & 12-Apr-07 & 10:31:50 & 1.1 & 6.15 &  20.06 (0.92) & 68.77 (0.80) &102.29 (0.78) &116.80 (0.99) & 45.85 (1.02) &  5.21 (0.78) &  2.58 (1.21)\\
44 & 14-Apr-07 & 13:52:01 & 1.1 & 6.30 &  22.01 (0.92) & 64.60 (0.84) & 96.85 (0.87) &134.60 (1.06) & 62.22 (1.18) &  9.45 (0.99) &  9.62 (1.27)\\
45 & 14-Apr-07 & 23:43:23 & 1.5 & 6.20 &  21.74 (1.26) & 77.65 (0.96) &137.29 (0.88) &232.24 (0.94) &162.83 (0.77) & 23.37 (0.80) & 27.34 (1.15)\\
[.3ex]\cline{4-12}\\[-1.6ex]
 - &   Average &       -  & 1.3 & 6.80 &  21.18 (1.15) & 66.09 (0.99) &107.77 (0.95) &162.59 (1.09) & 95.33 (0.99) & 14.39 (0.87) & 18.68 (0.95)\\
\enddata
\label{tab4}
\end{deluxetable}
\begin{figure}[ht]
\centering
\includegraphics[width=1.00\linewidth]{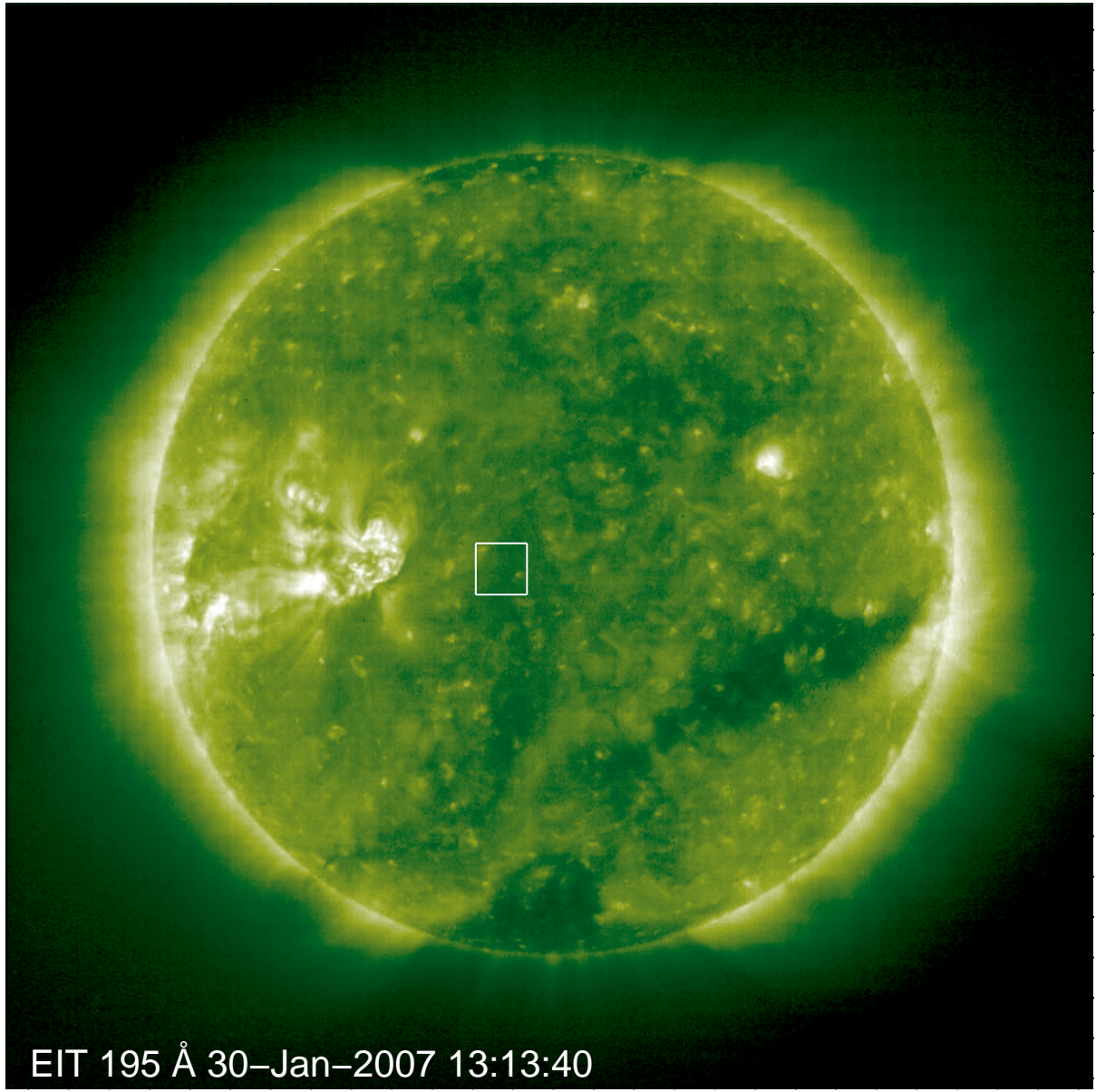}
\caption{Full-disk EIT 195\,\AA\, image showing the field-of-view of the EIS study analyzed.
\label{fig1}}
\end{figure}
\begin{figure}[ht]
\centering
\includegraphics[width=1.00\linewidth]{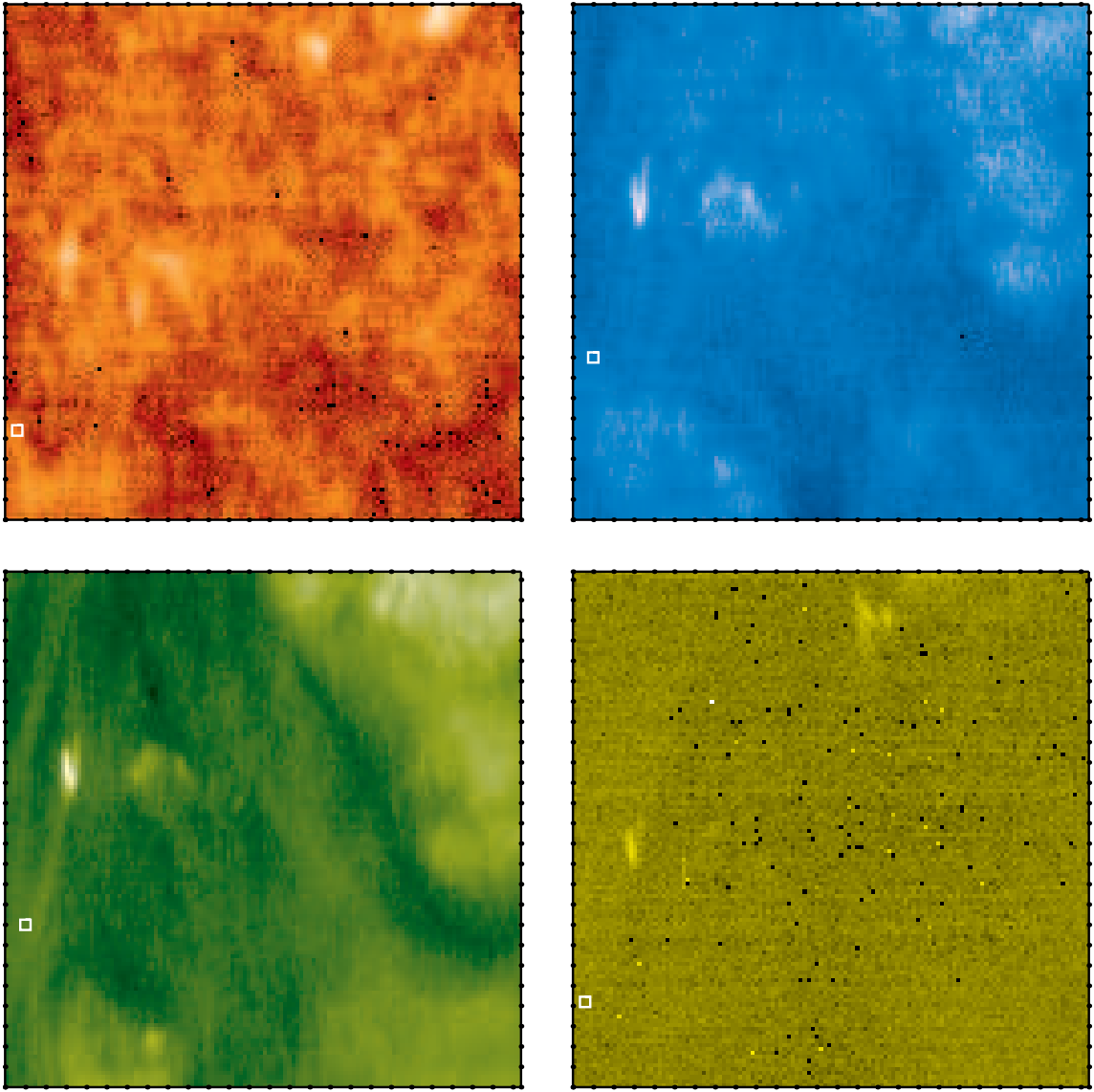}
\caption{EIS images of the observed quiet Sun region. Top Left:
\ion{Si}{7} 275.352\,\AA. 
Top Right: \ion{Fe}{10} 184.537\,\AA. 
Bottom Left: \ion{Fe}{12} 195.119\,\AA. 
Bottom Right: \ion{Fe}{15} 284.160\,\AA. The single pixel used in \S \ref{demsinglepix} 
is shown as a small (not to scale) box on each image.
\label{fig2}}
\end{figure}
\begin{figure}[ht]
\centering
\includegraphics[width=1.00\linewidth,viewport=-15 0 360 360,clip]{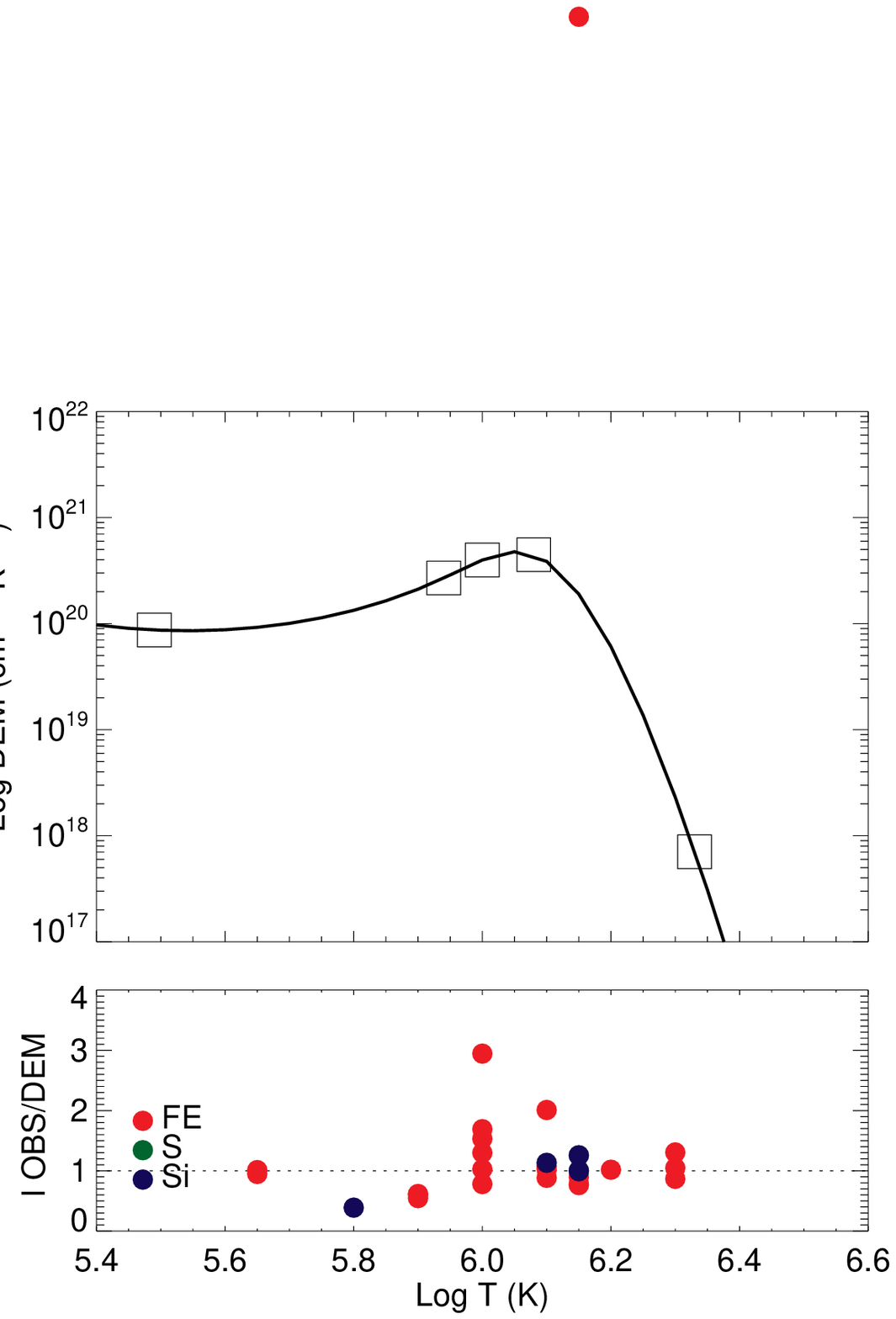}
\caption{The best case DEM solution for the spectral lines shown in Table \ref{tab1} plotted
as a function of temperature (upper panel). The squares show the positions of the 
spline knots. The lower panel shows ratios of the observed to
DEM predicted intensities. The coronal abundance of \citet{feldman_etal1992} for Fe has been
reduced by $\sim$ 40\% for this calculation.}
\label{fig3}
\end{figure}
\begin{figure}[ht]
\centering
\includegraphics[width=1.00\linewidth]{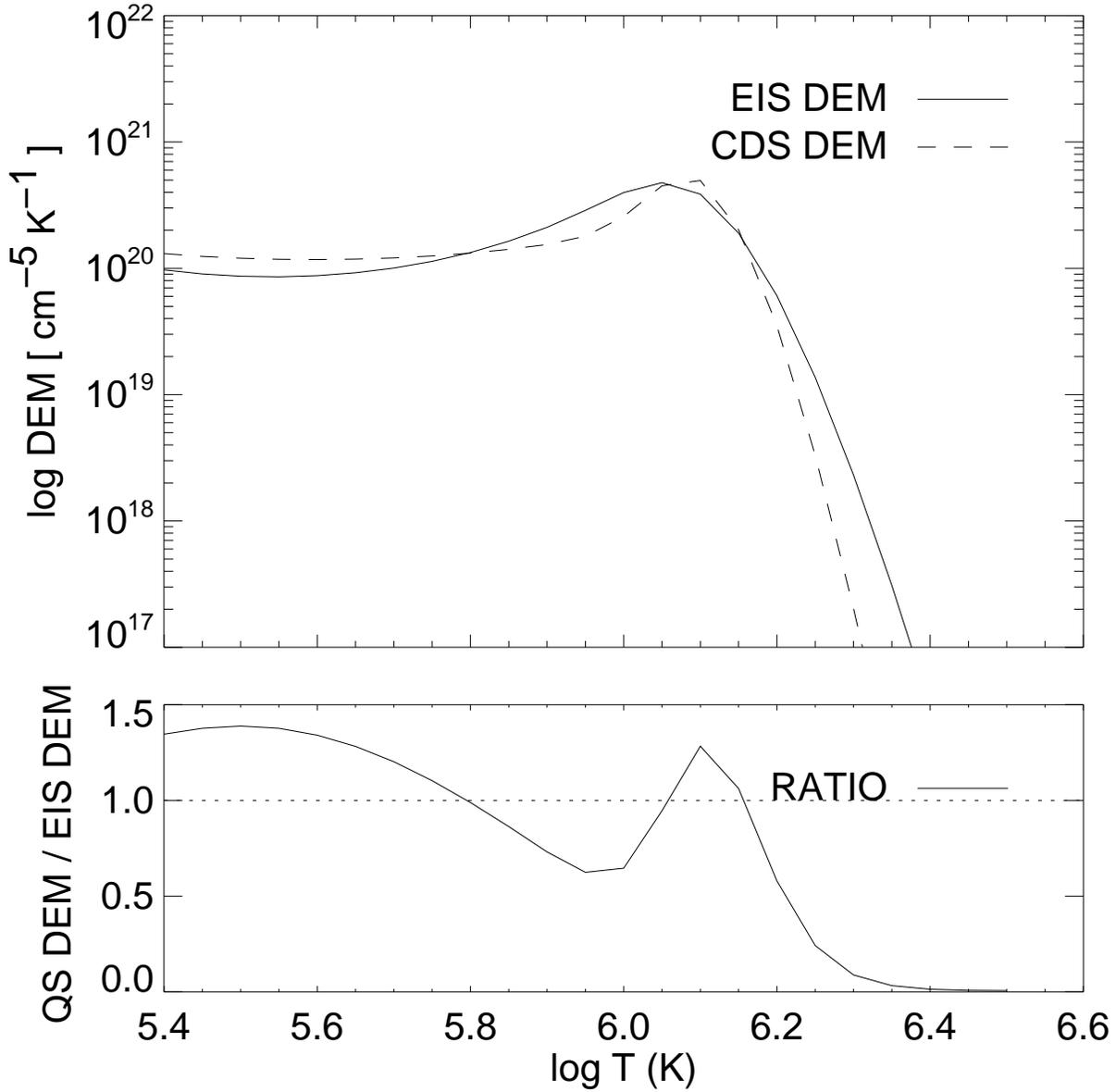}
\caption{Comparison between the EIS best case DEM and the DEM for the quiet corona
derived by \citet{brooks&warren_2006}. 
The Ratio of the previous DEM to the EIS DEM is shown in the lower panel.}
\label{fig4}
\end{figure}
\begin{figure}[ht]
\centering
\includegraphics[width=1.00\linewidth,viewport=-15 0 360 360,clip]{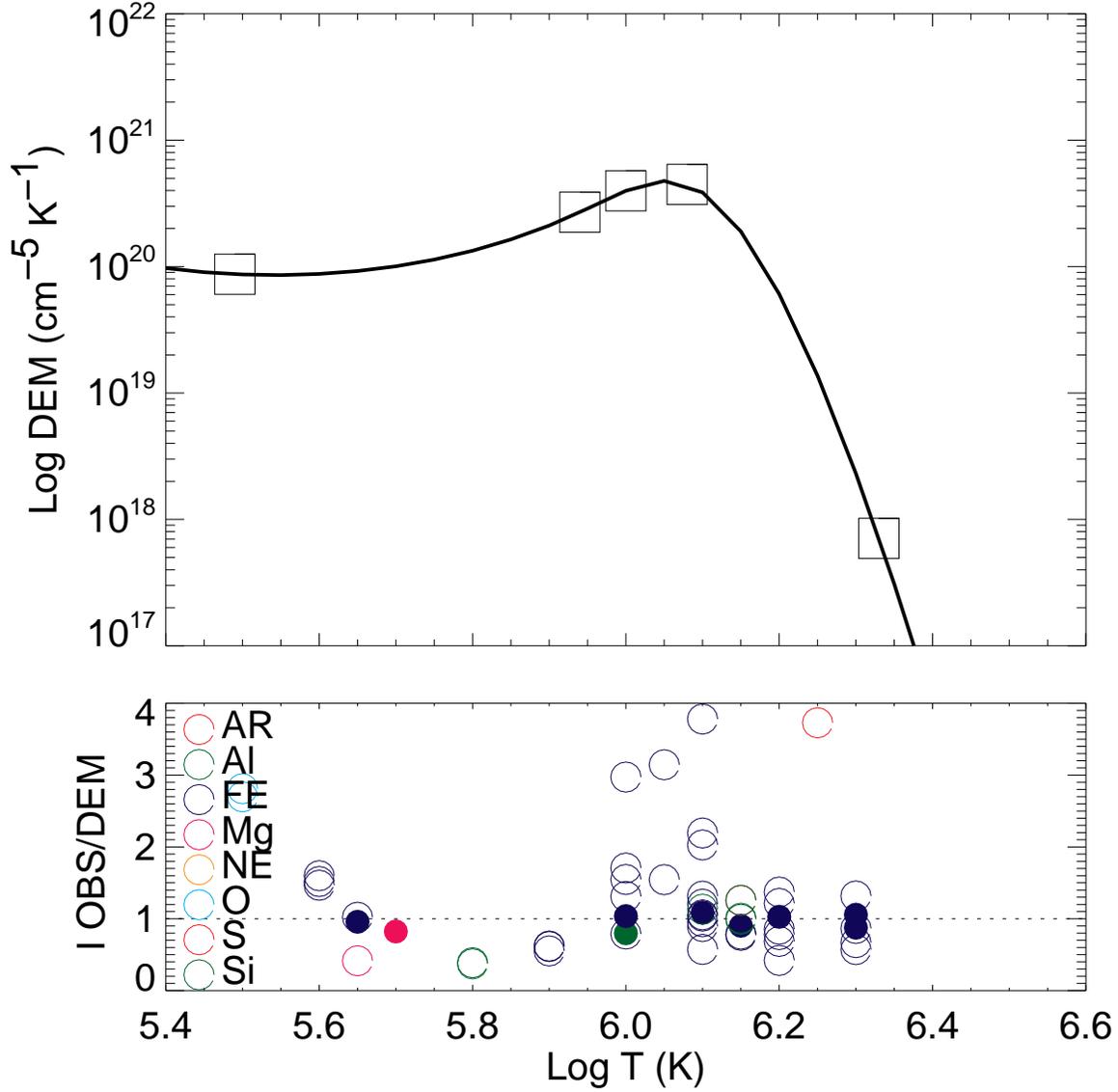}
\caption{The DEM solution for all the spectral lines studied in this paper plotted
as a function of temperature (upper panel). The squares show the positions of the 
spline knots. The lower panel shows ratios of the observed to
DEM predicted intensities. Filled circles represent the lines retained to produce the
subset and empty circles represent the lines that were deselected because there was a 
better candidate line available at the same temperature.}
\label{fig5}
\end{figure}
%
\begin{figure}[ht]
\centering
\includegraphics[width=1.00\linewidth,clip]{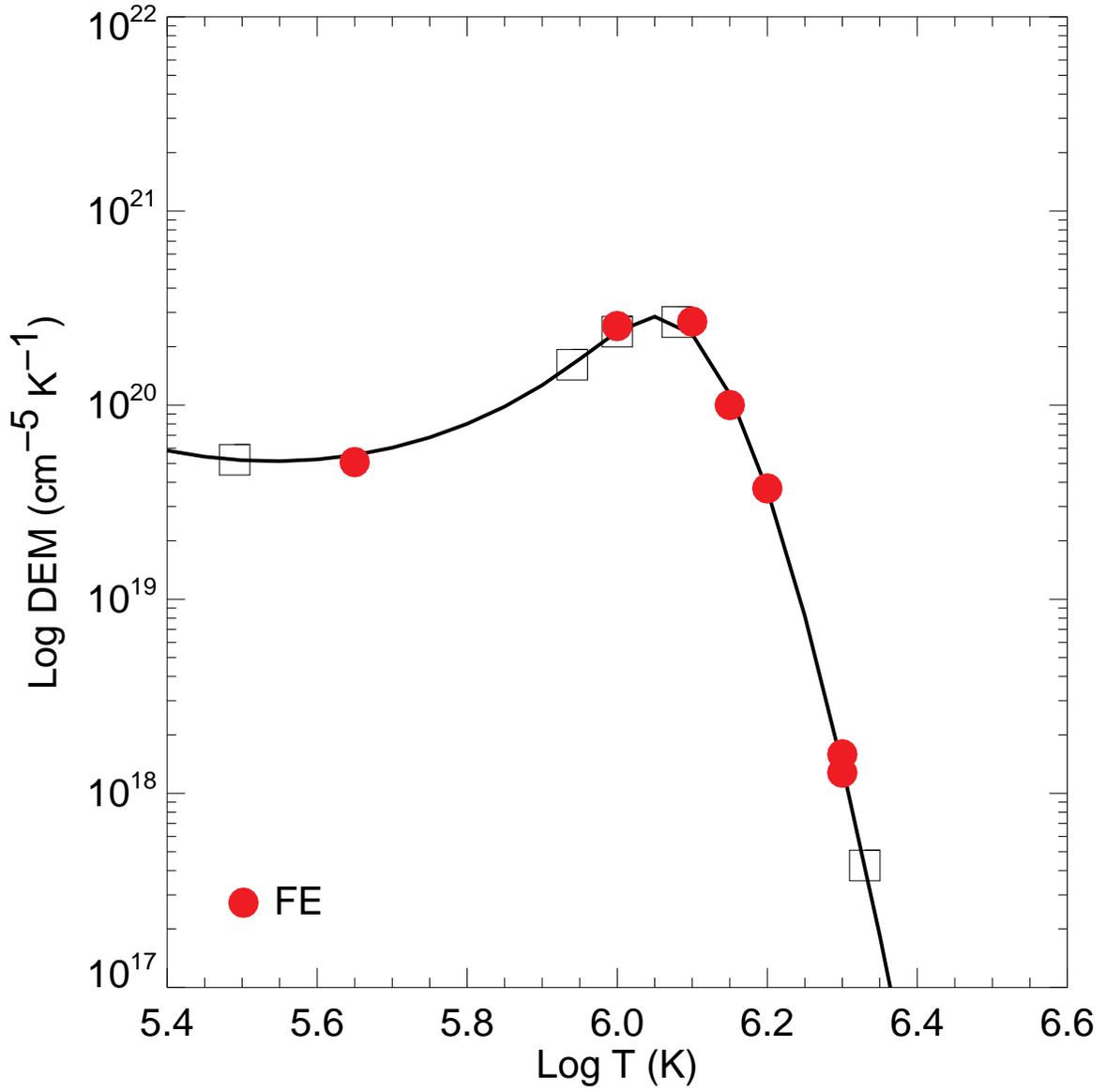}
\caption{DEM distribution derived from a single EIS pixel.
\label{fig6}}
\end{figure}
\begin{figure}[ht]
\centering
\includegraphics[width=1.00\linewidth,clip]{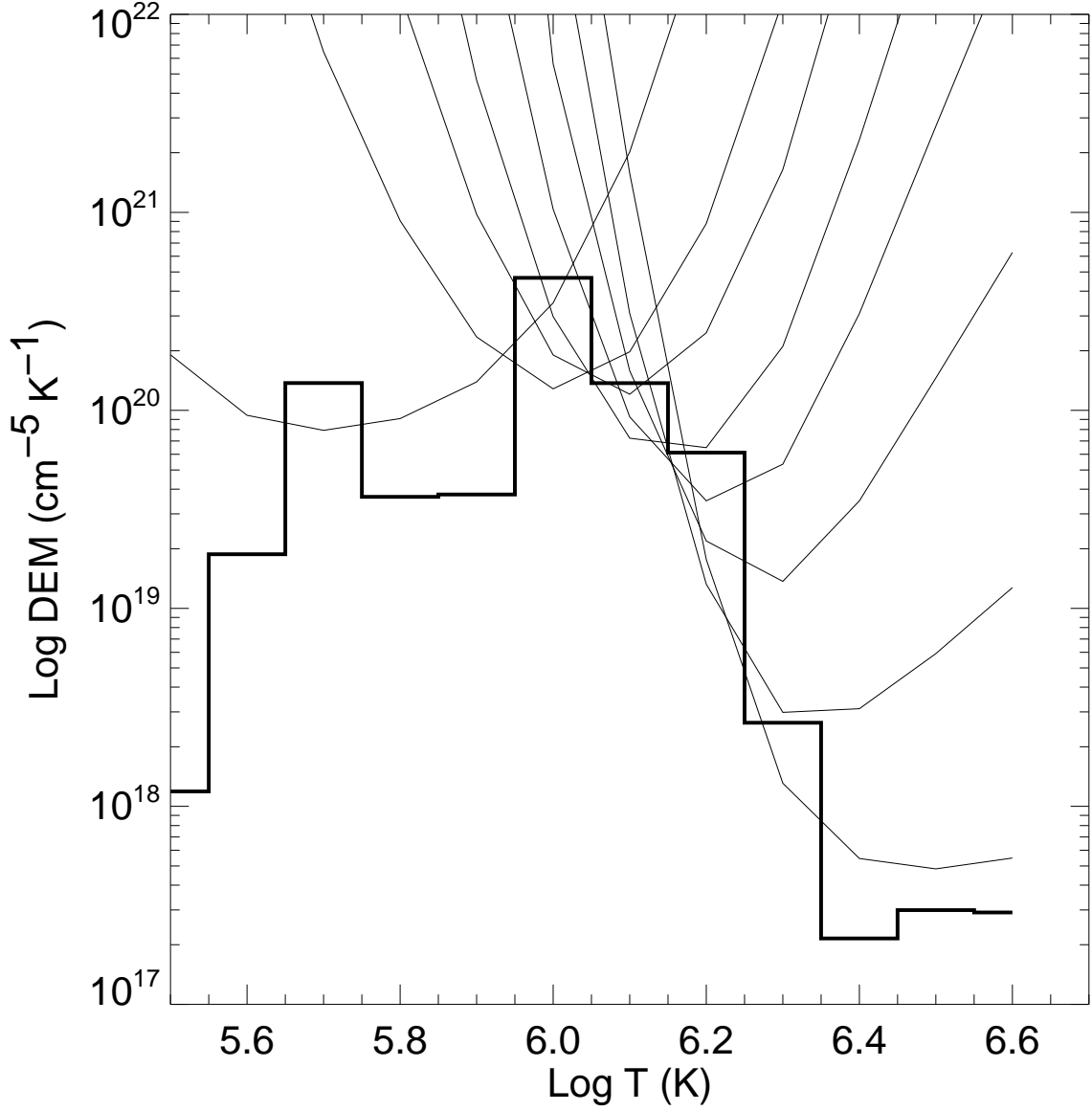}
\caption{DEM distribution (histogram) derived from a single EIS pixel using the MCMC algorithm. DEM
loci curves calculated for each line are also plotted.
\label{fig7}}
\end{figure}
\begin{figure}[ht]
\centering
\includegraphics[width=1.00\linewidth]{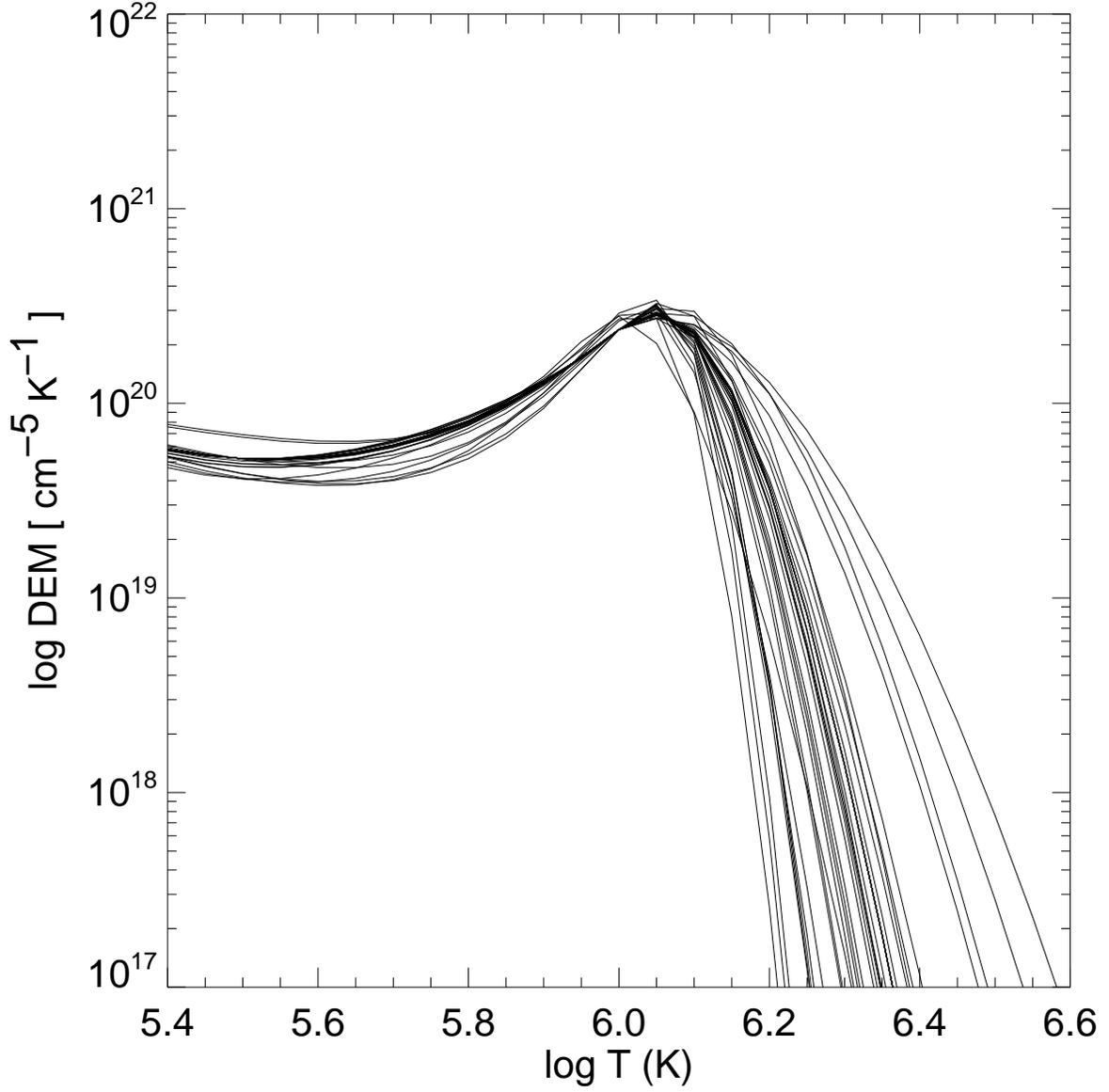}
\caption{DEM solutions for all the datasets as a function of temperature. Note the similarity in shape.
\label{fig8}}
\end{figure}
%
\end{document}